\documentclass[a4paper,11pt,fleqn]{article}
\usepackage{amsfonts,latexsym}

\def\nn{\nonumber}
\def\non{\nonumber\\}
\def\be{\begin{equation}}
\def\ee{\end{equation}}
\def\ben{\begin{displaymath}}
\def\een{\end{displaymath}}
\def\ba{\begin{eqnarray}}
\def\ea{\end{eqnarray}}

\def\d{\delta}
\def\e{\varepsilon}
\def\f{\varphi}

\def\g{\gamma}

\def\m{\mu}

\def\O{\Omega}

\def\s{\sigma}

\def\e{\epsilon}

\def\cL{{\cal L}}

\def\cV{{\cal V}}

\def\R{\mathbb{R}}

\def\la{\label}
\def\c{\cite}

\def\Ref#1{(\ref{#1})}

\def\f{\frac}

\def\i{\infty}
\def\p{\partial}

\def\tr{{\rm tr}}

\def\coset{SL(2,\R)/SO(2)}

\def\gg{\mathfrak{g}}
\def\gh{\mathfrak{h}}
\def\gk{\mathfrak{k}}
\def\vh{{\hat{\cV}}}

\begin{document}
\renewcommand{\thefootnote}{\fnsymbol{footnote}}
\begin{flushright}
DESY 96-245\\
gr-qc/9611061\\
November 1996\\
\vspace*{0.7cm}
\end{flushright}
\begin{center}
{\LARGE  Poisson Realization}\medskip\\
{\LARGE and Quantization of the Geroch Group}\bigskip\\ 
{\large D. Korotkin\footnote{On leave of absence from Steklov
Mathematical Institute, Fontanka, 27, St.Petersburg 191011 Russia} and
H. Samtleben\bigskip\medskip\\ { II. Institut f\"ur Theoretische
Physik, Universit\"at Hamburg,}\\ { Luruper Chaussee 149, 22761
Hamburg, Germany}}\\ \small E-mail: korotkin@x4u2.desy.de,
jahsamt@x4u2.desy.de
\end{center}
\renewcommand{\thefootnote}{\arabic{footnote}}
\setcounter{footnote}{0}
\vspace*{0.9cm}
\hrule
\begin{abstract}
\noindent
The conserved nonlocal charges generating the Geroch group with
respect to the canonical Poisson structure of the Ernst equation are
found. They are shown to build a {\em quadratic} Poisson algebra,
which suggests to identify the quantum Geroch algebra with Yangian
structures.
\end{abstract}
\medskip

\hrule
\vspace*{1.2cm}

Geroch's discovery of an infinite-dimensional group of symmetries
\c{Gero71b} was one of the essential steps in the study of Einstein's
equations with two Killing vector fields. The understanding of the
algebraic structure underlying the Geroch group essentially improved
with the construction of the corresponding linear system
\c{BelZak78,Mais78}. Closer analysis of this system further provided
different insightful realizations \c{Juli83,ChaGe88,BreMai87} of the
symmetry algebra --- meanwhile identified as half of an affine
Kac-Moody algebra.

In this letter, the Poisson realization of the Geroch group that has
been missed so far is revealed. We identify the infinite set of
conserved charges that generate the infinitesimal symmetry
transformations with respect to the canonical Poisson structure. For
the quantum theory, where the classical algebra of symmetries turns
into a spectrum-generating algebra, the Poisson realization is of
essential importance, since it is exactly the quantum pendant of this
algebra according to the representations of which the spectrum of
physical states is classified.

Thus, indeed surprising is the further obtained result \Ref{Q}, that
in spite of the fact that the Geroch algebra of transformations itself
is known to be linear, the Poisson algebra of associated charges comes
out to be quadratic. Its action however linearizes being restricted to
the subset of solutions non-singular at $t\!=\!0$, in which all the
Geroch charges vanish. The natural quantum counterpart of the
quadratic algebra is the affine quantum group (Yangian), further
related to knot theory. This indicates the link to arising of these
structures in the framework of the loop variable approach to quantum
gravity \c{RovSmo88,Asht91}.

Let us begin by fixing the notation, mainly following \c{Nico91}. The
model of symmetry reduced $4D$ Einstein gravity is embedded into the
general framework of $2D$ dilaton-gravity coupled coset $\s$-models,
which makes the underlying group theoretical structure more
transparent. As basic canonical variables we consider fields
$\cV(x^\m)$ mapping a $2D$ Lorentzian world-sheet parametrized by
coordinates $x^0$ and $x^1$ into a group $G$. The corresponding
currents $J_\m\!\equiv\!J_\m^at_a\!\equiv\!\cV^{-1}\p_\m\cV$ live in
the Lie algebra $\gg$ with basis generators $t_a$. Denote by
$\gh\!\subset\!\gg$ the Lie algebra underlying the maximal compact
subgroup $H\!\subset\!G$; then $\gg$ admits the decomposition into an
orthogonal sum $\gg=\gh\oplus\gk$. Accordingly, the physical currents
$J_\m$ are decomposed as
\be\la{PQ}
J_\m=\cV^{-1}\p_\m\cV\equiv Q_\m+P_\m\;;
\qquad\mbox{with}\quad Q_\m\in\gh\,,\enspace P_\m\in\gk\;.
\ee

We start from the $2D$ Lagrangian
\be\la{L}
\cL = {\textstyle \frac12}t\;\tr (P_0^2-P_1^2)\;,
\ee
that is obtained from dimensional reduction of general higher
dimensional gravities and supergravities leading to various coset
spaces $G/H$. In particular, Einstein gravity with two commuting
(spatial) Killing vector fields takes the form \Ref{L} with the field
$\cV$ parametrizing the coset $\coset$.

The model \Ref{L} differs from the well known chiral field model only
by appearance of the dilaton field $t$, which is remnant of the size
of the compactified dimensions of the higher-dimensional
space-time. From the treatment of the so-called Einstein-Rosen waves
(see the discussion in \c{Kuch71}) we further adopt the identification
of $t$ with one of the coordinates of the $2D$ worldsheet. For
definiteness we set $t\!\equiv\!x^0$, hence measuring time by the
determinant of the compactified part of the $4D$ metric. The
spatial coordinate is denoted by $x\!\equiv\!x^1$.

The form of the Lagrangian \Ref{L} reveals its invariance under rigid
transformations ${\rm g}\!\in\!G$ and local transformations ${\rm
h}(x,t)\!\in\!H$
\be\la{gaugeP}
\cV(x,t)\mapsto {\rm g}^{-1}\cV(x,t){\rm h}(x,t)\;,\quad 
P_\m\!\mapsto\!{\rm h}^{-1}P_\m{\rm h}\;,
\ee
that remain from original constant linear coordinate transformations
and local Lorentz transformations respectively. The equations of
motion derived from \Ref{L} read
\be\la{eqm}
D^\m(t\, P_\m) = D_0(t\, P_0) - D_1(t\, P_1) = 0\;,
\ee 
with the covariant derivative $D_\m P_\nu\!\equiv\!\p_\m P_\nu
+[Q_\m\,,P_\nu]$. In terms of the field $g\!\equiv\!\cV\cV^t
\in G$, this becomes the more familiar form of the Ernst equation
\c{Erns68}
\be\la{Ernst}
\p^\m(t\, g^{-1}\p_\m g) ~=~ \p_0(t\, g^{-1}\p_0 g) 
- \p_1(t\, g^{-1}\p_1 g) ~=~ 0 \;.
\ee

We now put up the canonical framework. Treating $Q_1$ and $P_1$ as
canonical variables, the corresponding momenta $\pi_Q$ and $\pi_P$ are
introduced together with the Poisson structure
\be\la{PB}
\left\{Q_1^a(x),\pi_Q^b(y)\right\} 
= \left\{P_1^a(x),\pi_P^b(y)\right\} 
= \d^{ab}\,\d(x\!-y)
\ee
at equal time $t\!\not=\!0$ \c{t0}. For convenience in later
calculations we switch to the index-free tensor notation. Denote
$\stackrel{1}{X}\equiv X\!\otimes I$ and $\stackrel{2}{X}\equiv
I\otimes X$. Let further $\O_\gg\!\equiv\!  t_a\!\otimes t^a$ be the
Casimir element of $\gg$, which allows the decomposition $\O_\gg
=\O_\gh+\O_\gk$ (since the direct sum $\gg=\gh\oplus\gk$ is orthogonal
with respect to the Cartan-Killing form).  The canonical brackets
\Ref{PB} in the tensor notation become
\ba
\left\{\stackrel1{Q}_1\!(x)\;,\,\stackrel2{\pi}_Q\!(y)\right\} 
&=& \O_\gh\, \d(x\!-y)\;,\la{PB1}\\
\left\{\stackrel1{P}_1\!(x)\;,\,\stackrel2{\pi}_P\!(y)\right\} 
&=& \O_\gk\, \d(x\!-y) \;.\nn
\ea
We state the fundamental non-vanishing Poisson brackets for the
physical fields, calculated from \Ref{PB1}:
\ba
\left\{\stackrel1{P}_0\!(x)\;,\,\stackrel2{Q}_1\!(y)\right\} &=&
- \frac2{t}\;\left[\,\O_\gh\;,\,
\stackrel1{P}_1\!(x)\,\right]\d(x\!-y) 
\;,\la{PBf}\\
\left\{\stackrel1{P}_0\!(x)\;,\,\stackrel2{P}_1\!(y)\right\} &=&
- \frac2{t}\;\left[\,\O_\gk\;,\,
\stackrel1{Q}_1\!(x)\,\right]\d(x\!-y) 
+ \frac2{t}\;\O_\gk\;\p_1\d(x\!-y)\;. \nn
\ea
This completes the basics of the canonical setting. Our next goal is
the identification of a family of conserved charges. As essential
tool, we canonically employ the linear system of the model
\c{BelZak78,Mais78}, that encodes the equations of motion \Ref{eqm}:
\be\la{ls}
\p_\m\vh(x,t,\g) = \vh(x,t,\g)L_\m(x,t,\g)\;;
\ee
for a group-valued function $\vh(x,\g)$ with spectral parameter $\g$
and
\be\la{LL}
L_\m\equiv Q_\m + \frac{1+\g^2}{1-\g^2}P_\m + 
\frac{2\g}{1-\g^2}\e_{\m\nu} P^{\nu} \;.
\ee
Compatibility of this linear system is equivalent to \Ref{eqm} if the
spectral parameter $\g$ has an explicit coordinate dependence
according to
\be\la{gamma}
\g(x,t,w) = -{\textstyle\frac1{t}}
\Big(w-x\pm\sqrt{(w-x)^2-t^2}\Big)\;.
\ee
The free integration constant $w$ in \Ref{gamma} may be understood as
the hidden ``constant spectral parameter'' of the linear
system. Actually the coordinate dependence of the spectral parameter
current is the essential difference between the gravity models and the
chiral field model; the linear system \Ref{ls} with constant $\g$
leads to equations of motion $D^\m P_\m\!=\!0$, which differ from
\Ref{eqm} exactly by the absence of the factor $t$.

With the linear system \Ref{ls} at hand, we define for fixed $t$ the
associated transition matrices $T(x,y,w)$ \c{FadTak87} by
\be\la{T}
T(x,y,w)\equiv \vh^{-1}(x,\g(x,w))\vh(y,\g(y,w))\;.
\ee
They obey
\ben
\p_tT(x,y,w) = -L_0(x,\g(x,w))T(x,y,w) + T(x,y,w)L_0(y,\g(y,w))\;.
\een
This already shows, that assuming asymptotically flat solutions, for
which the currents $Q_\m, P_\m$ and hence $L_\m$ vanish at spatial
infinity, the transition matrices $T(w)\equiv
T(x\!\rightarrow\!-\i,y\!\rightarrow\!\i,w)$ become conserved
quantities:
\be
\p_tT(w)=0\;.
\ee
Thus, we obtain a family of time-independent quantities in complete
analogy to the chiral field model \c{Husa96}. Expanding $T(w)$
around $w\!=\!\infty$ according to
\be\la{exp}
T(w) = I + \frac1{w}T_1 + \frac1{w^2}T_2 + \dots
\ee
allows to extract a countable set of conserved charges. For
illustration, we state the first two expressions, which may explicitly
checked to be time-independent due to the equations of motion
\Ref{eqm}:
\ba
T_1 &=& 
t\int_{-\infty}^{\infty}{\rm d}x\;\cV(x)P_0(x)\cV^{-1}(x)  
\;,\la{charges}\\
T_2 &=& t^2\int_{-\infty}^{\infty}{\rm d}x\;
\cV(x)P_0(x)\cV^{-1}(x)
\int_{x}^{\infty}{\rm d}y\;\cV(y)P_0(y)\cV^{-1}(y) \non
&&{}+ {\textstyle \frac12} t^2\int_{-\infty}^{\infty}{\rm d}x\;
\cV(x)P_1(x)\cV^{-1}(x) \non
&&{}+ t\int_{-\infty}^{\infty}{\rm d}x\; 
x\cV(x)P_0(x)\cV^{-1}(x)\;. \nn
\ea

We can further calculate the Poisson structure of the transition
matrices $T(w)$, tracing back their dependence on the basic currents
via \Ref{T} and \Ref{LL}. Using standard techniques \c{FadTak87}, this
is a more or less straightforward although lengthy exercise.  The
explicit coordinate dependence of the spectral parameter makes parts
of the calculation more tedious, however, it is exactly this
dependence, that finally ensures the result to be the well-defined and
surprisingly simple expression
\be\la{Q}
\left\{\,\stackrel1{T}\!(v)\;,\;\stackrel2{T}\!(w)\,\right\} =
\left[\,\frac{2\O_\gg}{v-w}\;,\;
\stackrel1{T}\!(v)\!\stackrel2{T}\!(w)\,\right]\;,
\ee
which is the first main result of this letter. In contrast to this
rather appealing quadratic structure, the analogous calculation for
the related chiral field model was spoiled by severe and incurable
ambiguities, that were caused by the non-ultralocal terms (containing
derivatives of delta-functions) in the basic Poisson brackets
\Ref{PBf} \c{VeEiMa84}. Although the same terms arise here, their
behavior is roughly speaking compensated by the coordinate dependence
of the spectral parameter \Ref{gamma}.

It is crucial to notice that all the charges contained in $T(w)$
become invisible
\be\la{van}
T(w)=I\;,
\ee
if the physical currents $Q_\m, P_\m$ are regular at $t\!=\!0$. This
can already be seen from \Ref{charges}: all the $T_m$ are independent
of $t$ and vanish for $t\!=\!0$. We should emphasize, that \Ref{van}
does not arrive as a constraint, but is understood in the way, that
the non-singular solutions live in the sector where \Ref{van} holds
(just as \Ref{Q} is an algebra of symmetries rather than a gauge
algebra of first-class constraints).

Despite this fact, which is another essential and surprising
difference in comparison with the chiral field model, the charges play
a rather nontrivial role in the non-singular subsector. Being
conserved quantities, they commute with the full (covariantized)
Hamiltonian and thus generate symmetries of the model. We explicitly
give the action of $T(w)$ on the physical currents up to $H$-gauge
transformations \Ref{gaugeP}:
\be\la{sym}
\left\{\,\stackrel1{T}\!(w)\;,\,
\stackrel2{P}_\pm\!(x)\,\right\}\; ~=~
\mp\frac{8\g^2\stackrel1{T}_-\!(x,w)\;\O_\gk\,
\stackrel1{T}_+\!(x,w)}{t^2(1\pm\g)^2(1-\g^2)}
 \;,
\ee
where we have defined
\ben
P_\pm\equiv P_0\pm P_1\;,\quad 
T_-(x,w)\equiv T(-\i,x,w)\;,\quad T_+(x,w)\equiv T(x,\i,w)\;.
\een
According to the expansion \Ref{exp}, we obtain in particular 
\ba
\Big\{\tr (JT_1),P_\pm(x)\Big\} &=& 0\;,\\
\Big\{\tr (JT_2),P_\pm(x)\Big\} &=& 
\mp2 \Big(\cV^{-1}(x)J\cV(x)\Big)\Big|_\gk\;\;;
\la{g2}
\ea
for a constant matrix $J$, where $|_\gk$ denotes the projection of
$\gg$ onto the coset subpart $\gk$.  Comparing this with the
infinitesimal action of the Geroch group as it has been made explicit
in \c{Nico91}, it turns out, that \Ref{g2} exactly coincides with the
action of the element $g_1$ from the affine algebra underlying the
Geroch group with commutation relations
\be\la{G}
\Big[\,g_m^a\,,\,g_n^b\,\Big] = f^{ab}_{~~c}\:g^c_{m+n}\;. 
\ee
In particular, the set of $g_1^a$ generates that half of the affine
algebra that acts nontrivially on the physical fields. Thus, having
identified the action of $g_1$ among the transformations \Ref{sym},
the action of the entire nontrivial half of the Geroch group is
enclosed in the $T(w)$.

{}From the mathematical point of view it may at first sight cause a
little surprise that the quadratic algebra \Ref{Q} contains half of an
affine algebra. However, this is actually nothing but an amusing
consequence of the vanishing of the charges in the non-singular
subsector, on which the Geroch group commonly has been
implemented. Recalling \Ref{van} it becomes clear, that in this sector
the matrix entries of $(T\!-\!I)^2$ Poisson-commute with all the
fields. Thus without changing the algebraic structure of the symmetry
generators, the relations \Ref{Q} are equivalent to:
\ba
\left\{\,\stackrel1{T}\!(v)\;,\;\stackrel2{T}\!(w)\,\right\} &=&
\left[\,\frac{2\O_\gg}{v-w}\;,\;
\stackrel1{T}\!(v)\!\stackrel2{T}\!(w) \,\right] \\
&& {}-\left[\,\frac{2\O_\gg}{v-w}\;,\;
\Big(\stackrel1{T}\!(v)-\stackrel1I\Big)
\Big(\stackrel2{T}\!(w)-\stackrel2I\Big)
\,\right] \non
&=& \left[\,\frac{2\O_\gg}{v-w}\;,\; 
\stackrel1{T}\!(v)+\stackrel2{T}\!(w)\,\right], \nn
\ea
which just describes half of an affine algebra \c{FadTak87}. In this
sense we may identify the $T_m$ from \Ref{exp} with the $g_{m-1}$ from
\Ref{G}. 

The Poisson realization of the Geroch group now opens up the way to
approach its quantum counterpart. The consistent quantization of the
quadratic structure \Ref{Q} is known \c{Fadd84,ChaPre94} and leads to
exchange relations of the Yangian type
\be\la{Yang}
R(u-v)\stackrel1{T}\!(u)\stackrel2{T}\!(v)
\,=\;\stackrel2{T}\!(v)\stackrel1{T}\!(u)R(u-v)\;,
\ee
where the matrix entries of $T(w)$ are now represented as operators on
a Hilbert space and
\be
R(u) = I+\f{2i\hbar\O_\gg}{u}
\ee
is a quantum $R$-matrix satisfying the Yang-Baxter equation with
spectral parameter. These quantum group structures are known to be
naturally linked to knot theory \c{ChaPre94} which in turn arises in
the loop variable approach to $4D$ quantum gravity
\c{RovSmo88,Asht91}. This suggests that quantization of the
midi-superspace model of dimensionally reduced gravity \Ref{L} already
reveals essential mathematical structures of the full quantum theory.

Since the quantum Geroch algebra \Ref{Yang} underlies an algebra of
classical symmetries \Ref{Q}, the spectrum of physical states will
admit decomposition into multiplets according to representations of
\Ref{Yang}. Although the complete related quantum model is still
missing, Ref.~\c{KorNic96} provides an exact quantization of the
isomonodromic truncation of \Ref{L} in a covariant ``two-time''
framework. It would be highly interesting to recover traces of the
mentioned multiplet structures among the exact solutions of the
Wheeler-DeWitt equation.

A more detailed presentation of the results stated in this letter will
appear elsewhere. We would like to thank H.~Nicolai for helpful
discussions. The work of D.~K. was supported by DFG Contract
Ni~290/5-1; H.~S. thanks Studienstiftung des Deutschen Volkes for
support.


\begin{thebibliography}{10}

\bibitem{Gero71b}
R.~Geroch, {J.~Math.~Phys.} {\bf 13}, 394 (1972).

\bibitem{BelZak78}
V.~Belinskii and V.~Zakharov, {Sov.~Phys.~JETP} {\bf 48}, 985 (1978).

\bibitem{Mais78}
D.~Maison, {Phys.~Rev.~Lett.} {\bf 41}, 521 (1978).

\bibitem{Juli83}
B.~Julia, in {\em Unified field theories in more than 4 dimensions},
  edited by V.~D.~Sabbata and E.~Schmutzer (World Scientific,
  Singapore, 1983).

\bibitem{ChaGe88}
L.-L.~Chau and M.-L.~Ge, {J.~Math.~Phys.} {\bf 30}, 166 (1988).

\bibitem{BreMai87}
P.~Breitenlohner and D.~Maison,
{Ann.~Inst.~H.~Poincar{\'e}.~Phys.~Th{\'e}or.} {\bf 46}, 215 (1987).

\bibitem{RovSmo88}
C.~Rovelli and L.~Smolin, {Phys.~Rev.~Lett.} {\bf 61}, 1155 (1988).

\bibitem{Asht91}
A.~Ashtekar, {\em Lectures on Nonperturbative Canonical Gravity}
  (World Scientific, Singapore, 1991).

\bibitem{Nico91}
H.~Nicolai, in {\em Recent Aspects of Quantum Fields}, edited by
  H.~Mitter and H.~Gausterer (Springer Verlag, Berlin, 1991).

\bibitem{Kuch71}
K.~Kucha\v{r}, {Phys.~Rev.} {\bf D4}, 955  (1971).

\bibitem{Erns68}
F.~Ernst, {Phys.~Rev.} {\bf 167}, 1175 (1968).

\bibitem{t0}
At $t\!=\!0$ the Lagrangian \Ref{L} of course becomes meaningless,
e.g.~visible in the singularity of the Poisson structure
\Ref{PBf}. For the conserved charges to be identified in the sequel
on the other hand, the limit $t\!\rightarrow\!0$ may be consistently
treated.

\bibitem{FadTak87}
L.~Faddeev and L.~Takhtajan, {\em Hamiltonian Methods in the Theory of
  Solitons} (Springer Verlag, Berlin, 1987).

\bibitem{Husa96}
It might be interesting to study the relation to a set of quantities
found in V.~Husain, {Phys.~Rev.} {\bf D53}, 4327 (1996), that
still exhibit {\em explicit}~ time-dependence and thus Poisson-commute
with the reduced Hamiltonian rather than with the full (covariantized)
one.

\bibitem{VeEiMa84}
H.~de~Vega, H.~Eichenherr, and J.~Maillet, {Commun.~Math.~Phys.}
  {\bf 92}, 507 (1984).

\bibitem{Fadd84}
L.~Faddeev, in {\em Les Houches, Session XXXIX, recent advances in
  field theory and statistical mechanics}, edited by J.-B.~Zuber and
  R.~Stora (North-Holland, Amsterdam, 1984).

\bibitem{ChaPre94}
V.~Chari and A.~Pressley, {\em A Guide to Quantum Groups} (Cambridge
  University Press, Cambridge, 1994).

\bibitem{KorNic96}
D.~Korotkin and H.~Nicolai, {Nucl.~Phys.} {\bf B475}, 397 (1996).

\end{thebibliography}

\end{document}